\documentclass[12pt]{article}
\usepackage{graphicx,amsmath,amssymb,amsfonts,color}
\usepackage{citesort,epsfig}

\input{text/intrinsic.pre}
\input{text/intrinsic.fig}

\begin{document}


\begin{titlepage}

\begin{tabular}{l}
\noindent\DATE
\end{tabular}
\hfill
\begin{tabular}{l}
\PPrtNo
\end{tabular}

\vspace{1cm}

\begin{center}
\renewcommand{\thefootnote}{\fnsymbol{footnote}}
{
\Large \TITLE
}

\vspace{1.25cm}
{\large  \AUTHORS}

\vspace{1.25cm}

\INST
\end{center}

\vfill

\ABSTRACT                 

\vfill

\end{titlepage}


\section{Introduction}
\label{sec:introduction}

Interactions of hadrons at high energy---such as those probed in 
$\mathrm{ep}$ scattering at HERA, $\bar{\mathrm{p}}\mathrm{p}$ 
scattering at the Tevatron, and $\mathrm{pp}$
scattering at the forthcoming LHC---are to be understood in terms of 
the interactions of their quark and gluon constituents.  The parton 
distribution functions (PDFs) that describe the proton's quark and gluon 
content are therefore essential for testing the Standard Model and 
searching for New Physics.

The PDFs are functions $f_a(x,\mu)$ where $x$ is the fraction of the 
proton's momentum carried by parton species $a$ at scale $\mu$, in a 
frame where that momentum is large.  For small values of 
$\mu$, corresponding to long distance scales, the PDFs express 
nonperturbative physics that is beyond the scope of present
calculations from first principles in QCD (although some progress 
has been made using lattice methods \cite{lattice}). 
Current practice is instead to parametrize the PDFs at a scale $\mu_0$ 
that is large enough for ${f_a(x,\mu)}$ to be calculated from 
${f_a(x,\mu_0)}$ at all $\mu > \mu_0$ by perturbation theory.
The unknown functions $f_a(x,\mu_0)$ are determined empirically by 
adjusting their parameters to fit a large variety of data at 
$\mu > \mu_0$ in a ``QCD global analysis'' \cite{cteq6,mrst}. 

A number of important processes, including Higgs boson production in 
certain scenarios, are particularly sensitive to the bottom or charm 
quark distributions $f_c(x,\mu)$ and $f_b(x,\mu)$.  In the global analyses 
that have been carried out so far, it is assumed that the charm content 
of the proton is negligible at $\mu \! \sim \! m_c$, and similarly that 
bottom is negligible at $\mu \! \sim \! m_b$, so these heavy quark 
components arise only perturbatively through gluon splitting in the DGLAP 
evolution.  The global fits are not inconsistent with this assumption, 
but the data sets they are based on do not yet include experiments that 
are strongly sensitive to heavy quarks, so substantially larger 
$c$ or $b$ content cannot be ruled out.  Direct measurements of 
$c$ and $b$ production in deep inelastic scattering are also consistent 
with an entirely perturbative origin for heavy quark 
flavors \cite{F2ccbarbbbar}, but those experiments are not sensitive to 
heavy quarks at large $x$.

Meanwhile, in the light-cone Fock space picture \cite{LightCone},
it is natural to expect nonperturbative ``intrinsic'' heavy quark 
components in the proton wave function \cite{BrodskyHoyer,Brodsky}. 
Furthermore, $s$ and $\bar{s}$ quarks each carry $\sim \! 1\%$ of the 
proton momentum at 
$\mu_0 = 1.3 \, \mathrm{GeV}$ \cite{cteq6}, which implies that states 
made of $u u d s \bar{s}$, together with gluons, make up a significant 
component of the proton wave function.  By analogy, one would expect 
$u u d c \bar{c}$ and $u u d b \bar{b}$ also to be present---although 
the degree of suppression caused by greater off-shell distances 
is difficult to predict.  A suppression as mild as 
$\sim 1/m_c^2$ has been derived using a semiclassical approximation 
for the heavy quark fields \cite{Polyakov}.

An alternative way to describe the proton in light-cone Fock space is in 
terms of off-shell physical particles---the ``meson cloud'' 
picture \cite{Navarra,Melnitchoukccbar,SteffensMT}. 
Specifically, the two-body state 
$\overline{D}^0 \Lambda_c^+$, where $\overline{D}^0$ is a $u\bar{c}$ meson 
and $\Lambda_c^+$ is a $udc$ baryon, forms a natural low-mass component. 
This is the flavor $SU(4)$ analog of the $K^+ \Lambda^0$ component that is 
a natural source of strangeness, and in particular of 
$f_{s}(x,\mu) \ne f_{\bar{s}}(x,\mu)$ \cite{strangecloud}.
A charm contribution from the two-body state $p \, \jpsi$ is also possible. 

The light-cone view is not developed to a point where the normalization 
of $uudc\bar{c}$ and $uudb\bar{b}$ components can be calculated with 
any confidence, though estimates on the order of $1\%$ have been found 
for intrinsic charm (IC) using a meson cloud model \cite{Navarra}, the 
MIT bag model \cite{BagModel}, and an SU(4) quark model \cite{Song}.
However, we can use the picture to predict the $x$-dependence of the 
non-perturbative contribution. 
A central feature of the light-cone models is that heavy quarks appear 
mainly at large $x$, because their contribution to the off-shell 
distance is proportional to $(p_\perp^{\,2} + m^2)/x$, so the suppression 
of far off-shell configurations favors large $x$ when $m$ is large. 
We will show that this feature leads to similar predictions for the shape 
in $x$ from a wide variety of specific models. 

Using the rough consensus of the models as a guide to the shape of 
$x$-dependence for intrinsic charm and bottom, it will be possible to 
estimate their normalization from a limited set of data.  This will be 
carried out in a future publication.
When more complete data become available, such as jet measurements with 
$c$- and $b$-tagging (either inclusive jets or jets produced in association 
with $W$, $Z$, or $\gamma$), it will become possible to extract the 
$x$-dependence empirically.  It will then be interesting to see if the 
model predictions for the $x$-dependence are borne out.

Models in which the $u u d c \bar{c}$ Fock space component is considered 
directly are described in Sec.~\ref{sec:fivequark}.  Models based on 
low-mass meson+baryon pairs are described in Sec.~\ref{sec:MesonBaryon}.
The model results are compared with expectations for 
perturbatively generated heavy quarks in Sec.~\ref{sec:perturbative}.
The model results are compared with the light quark and gluon 
distributions in Sec.~\ref{sec:lightquark}.
Conclusions are summarized in Sec.~\ref{sec:conclusion}, where
simple parametrizations of all of the model predictions are tabulated 
for convenience in later work.
The connection between the light-cone description and ordinary Feynman 
diagrams, which is used in Secs.~\ref{sec:fivequark} 
and \ref{sec:MesonBaryon}, is derived in an Appendix.

\section{Five-quark models}
\label{sec:fivequark}

The probability distribution for the 5-quark state $u u d c \bar{c}$ 
in the light-cone description of the proton can be written as
\begin{equation}
dP = {\mathcal N} \, 
\prod_{j=1}^5 \! {{dx_j} \over {x_j}} \,
\delta(1 - \sum_{j=1}^5 x_j) \,
\prod_{j=1}^5 \! d^2 p_{j {\scriptscriptstyle\perp}} \,
\delta^{(2)}(\sum_{j=1}^5 p_{j {\scriptscriptstyle\perp}}) \,
{{F^2} \over {(s - m_0^{\,2})^2}} \; ,
\label{eq:fq1}
\end{equation}
where
\begin{equation}
s = 
\sum_{j=1}^{5} (p_{j {\scriptscriptstyle\perp}}^{\,2} \, + \, 
m_j^{\, 2}) / x_j 
\label{eq:fq2}
\end{equation}
and ${\mathcal N}$ is a normalization constant.  Eq.~(\ref{eq:fq1}) contains 
a wave function factor $F^2$ that characterizes the dynamics of the bound 
state. This factor must suppress contributions from large values of $s$ 
to make the integrated probability converge. 
An elementary derivation of Eq.~(\ref{eq:fq1}) is given in the 
Appendix. 

\subsection{The BHPS model}
\label{subsec:BHPSmodel}

A simple model for the $x$-dependence of charm can be obtained by neglecting 
the $p_\perp$ content, the $1/x_j$ factors, and $F^2$ in Eq.~(\ref{eq:fq1}). 
Further approximating the charm quark mass as large compared to all the other 
masses yields
\begin{equation}
dP \propto 
\prod_{j=1}^5 dx_j \, \delta(1 - \sum_{j=1}^5 x_j) \,
(1/x_4 + 1/x_5)^{-2}  \; ,
\label{eq:fq3}
\end{equation}
where $x_4 = x_c$ and $x_5 = x_{\bar{c}}$.
Carrying out all but one of the integrals and normalizing to an assumed total 
probability of $1\%$ yields 
\begin{equation}
{{dP} \over {dx}} =  f_c(x) = f_{\bar{c}}(x) = 
6\, x^2 \left[6 \, x \, (1+x) \, \ln x \, + \, 
(1-x) (1 + 10x + x^2) \right] \; ,
\label{eq:fq4}
\end{equation}
where $x = x_c$ or $x_{\bar{c}}\,$.  Equation~(\ref{eq:fq4}) was first 
derived by Brodsky, Hoyer, Peterson and Sakai \cite{BrodskyHoyer}, and 
has been used many times since.  We will use this BHPS model as a 
convenient reference for comparing all other models. 

Charm distributions that arise when the transverse momentum content 
of Eq.~(\ref{eq:fq1}) is not deleted are derived in the following 
subsections.

\figone

\subsection{Exponential suppression}
\label{subsec:Exponential}

A plausible conjecture would be that high-mass configurations in 
Eq.~(\ref{eq:fq1}) are suppressed by a factor 
\begin{equation}
  F^2 = e^{ - (s - m_0^{\,2}) / \Lambda^2} \; .
\label{eq:fq6}
\end{equation}
This exponential form makes the total probability integral converge for 
any number of constituents, while a power law would 
not (see Eq.~(\ref{eq:app10}) in the Appendix).

Figure~\ref{fig:fig1}(a) shows the charm distribution for 
several choices of the parameter $\Lambda$ in Eq.~(\ref{eq:fq6}).
The mass values used were 
$m_0 = 0.938 \, \mathrm{GeV}$ and
$m_4 = m_5 =  1.5 \, \mathrm{GeV}$.
Constituent quark masses $m_1 = m_2 = m_3 = m_0/3$ were 
used for the light quarks, but even setting those masses to zero
instead yields very similar results. 
All curves are normalized to $1\%$ integrated probability.
The results are qualitatively similar to the BHPS model, but are somewhat 
smaller in the region $x > 0.5$, which is the most important region 
as shown in Sec.~\ref{sec:lightquark}.

\subsection{Power-Law suppression}
\label{subsec:PowerLaw}

Alternatively, we might assume that high-mass five-quark states are 
suppressed only by a power law, say
\begin{equation}
F^2 \propto 
(s + \Lambda^{2})^{-n}
\; .
\label{eq:fq7}
\end{equation}
Figure~\ref{fig:fig1}(b) shows the charm distribution for several choices 
of the parameter $\Lambda$ with $n = 4$.  The results are again rather 
similar to the BHPS model, and again smaller than that model at large $x$. 
This behavior is fairly insensitive to the choice of $n$:  similar results 
were found for $n = 3$ (not shown),  while large values of $n$ revert to 
the exponential form of Section~\ref{subsec:Exponential}.  Values $n \le 2$ 
are unphysical, since they would make the total probability diverge.
The value $n=3$ is perhaps the most natural, since it leads to a
dependence $\sim 1/m_c^2$ that is in line with the result of \cite{Polyakov}.

\figtwo

\subsection{Quasi-Two-body suppression}
\label{subsec:Twobody}

Another approach to the suppression of high-mass Fock space components can 
be made on the basis of quasi-two-body states, such as those that will be 
considered explicitly in Section~\ref{sec:MesonBaryon}.  For instance, we 
might assume the relevant 5-quark configurations are grouped as 
$(u d c)(u \bar{c})$, in which case a plausible wave function factor would be
\begin{equation}
F^2 \propto 
       (s_{124} + \Lambda_{124}^{\;2})^{-2} \times 
       (s_{35}  + \Lambda_{35}^{\,2})^{-2}
\label{eq:fq8}
\end{equation}
where 
\begin{eqnarray}
s_{124} &=&
(p_{1 {\scriptscriptstyle\perp}}^{\,2} + m_1^{\,2})/x_1 \, + \, 
(p_{2 {\scriptscriptstyle\perp}}^{\,2} + m_2^{\,2})/x_2 \, + \, 
(p_{4 {\scriptscriptstyle\perp}}^{\,2} + m_4^{\,2})/x_4 \\
s_{35} &=& 
(p_{3 {\scriptscriptstyle\perp}}^{\,2} + m_3^{\,2})/x_3 \, + \, 
(p_{5 {\scriptscriptstyle\perp}}^{\,2} + m_5^{\,2})/x_5 \;.
\end{eqnarray}
Figure~\ref{fig:fig2} shows the $c$ and $\bar{c}$ distributions according 
to this assumption.  The parameter choices were 
$\Lambda_{124} = 2.5 \, \mathrm{GeV}$ and 
$\Lambda_{35} = 2.0 \, \mathrm{GeV}$, but other plausible choices lead 
to similar results.  Note that there is a small difference between the 
$c$ and $\bar{c}$ distributions in this model, 
with---perhaps surprisingly---$\bar{c}(x) > c(x)$ 
at large $x$.\footnote{%
One might have expected $c(x) > \bar{c}(x)$ because the $c$ quark comes 
from the heavier (baryonic) subgroup.  But in fact, the two subgroups 
share the proton momentum fairly equally, while the $\bar{c}$ retains 
more of the momentum of its (mesonic) subgroup because it shares 
that subgroup momentum only with a single quark.
\label{footnote:fn1}
}
Observing a $\bar{c}(x)\,$--$\, c(x)$ difference would of course 
definitively prove a non-perturbative origin for charm.

Alternatively, we might assume the relevant 5-quark configurations are 
grouped as $(u u d)(c \bar{c})$, so a plausible wave function factor 
would be
\begin{equation}
F^2 \propto 
       ( s_{123} + \Lambda_{123}^{\;2})^{-2} \times 
       (s_{45} + \Lambda_{45}^{\,2})^{-2}
\; .
\label{eq:fq9}
\end{equation}
The result of this assumption with $\Lambda_{123} = 1 \, \mathrm{GeV}$ 
and $\Lambda_{45} = 3 \, \mathrm{GeV}$ is also shown in 
Fig.~\ref{fig:fig2}.  It happens to be very similar to the average 
of $c$ and $\bar{c}$ from the preceding model.

\figthree
\figfour
\figfive

\section{Meson+Baryon models}
\label{sec:MesonBaryon}

Another way to picture the proton in light-cone Fock space is as a 
superposition of configurations of off-shell physical particles.  
In particular, the two-body state $\overline{D}^0 \Lambda_c^+$, 
where $\overline{D}^0$ is a $u\bar{c}$ meson and $\Lambda_c^+$ is a 
$udc$ baryon, is a natural low-mass component. 

We can model the $\overline{D}^0 \Lambda_c^+$ probability distribution 
in the proton using Eq.~(\ref{eq:app10}) with $N = 2$ and 
$F^2 \propto (s_{D\Lambda} + \Lambda_p^{\,2})^{-2}\,$.
The physical masses are 
$m_0 = 0.9383$, $m_1 = 1.8641$, and $m_2 = 2.2849$ in $\mathrm{GeV}$. 
We then model the $u d c$ distribution in $\Lambda_c^+$ similarly, using 
$N=3$ and $F^2 \propto (s_{udc} + \Lambda_{\Lambda}^{\,2})^{-2}$, with
$m_0 = 2.2849$, $m_1 = m_2 = 0.938/3$, and $m_3 = 1.6\,$. 
We similarly model the $u \bar{c}$ distribution in $\overline{D}^0$ using 
$N = 2$ and $F^2 \propto (s_{u\bar{c}} + \Lambda_D^{\,2})^{-2}$,  with 
$m_0 = 1.8641$ and $m_1 = m_2 = 1.60\,$.
(The charm quark mass here must be taken $> \! 1.55$ to keep 
$m_{\bar{c}} + m_u > m_{\overline{D}}$ for stability.)

The $c$ and $\bar{c}$ distributions in the proton follow from 
convolutions of the distributions defined above:
\begin{equation}
{{dP} \over {dx}} = 
\int_0^1 \! dx_1 \, f_1(x_1) \int_0^1 \! dx_2 \, f_2(x_2) \, 
\delta(x - x_1 x_2) =
\int_x^1{{dy} \over {y}} \, f_1(y) \, f_2(x/y) \; .
\label{eq:convolution}
\end{equation}
Figure~\ref{fig:fig3} shows the $c$ distribution from 
$p \to \overline{D}^0 \Lambda_c^+$. 
The result is very similar to the BHPS model.  Figure~\ref{fig:fig4} shows 
the $\bar{c}$ distribution from $p \to \overline{D}^0 \Lambda_c^+$. 
On comparing Figs.~\ref{fig:fig3} and \ref{fig:fig4}, we once again find 
$\bar{c}(x) > c(x)$ at large $x$---for the same reason as described in 
footnote \ref{footnote:fn1}.  This was observed previously in a 
meson cloud model that is rather similar to this one \cite{SteffensMT}. 
It is opposite to the asymmetry predicted by \cite{BrodskyMa}.

A contribution from the two-body state $p \, \jpsi$ is also possible. It is 
even slightly favored over $\overline{D}^0 \Lambda_c^+$ by having a lower 
threshold mass:
$m_p + m_{\jpsi} = 4.035 \, \mathrm{GeV} < 
m_D + m_{\Lambda_c} = 4.149 \, \mathrm{GeV}$.  (This is in contrast to the 
SU(4)-analog case of strangeness, where $K^+ \Lambda^0$ is strongly favored 
over $p \, \phi^0$ by 
$m_{K^+} + m_{\Lambda^0} = 1.609 \, \mathrm{GeV} \ll
m_p + m_\phi = 1.958 \, \mathrm{GeV}$.)
Figure~\ref{fig:fig5} shows the $c = \bar{c}$ distribution from the model of 
a $p \, \jpsi$ Fock space component and the model of $c$ or $\bar{c}$ in 
$\jpsi$.  A range of reasonable parameters for the suppression of large 
masses was tried and all give similar results.

\figsix
\figseven

\section{Comparison with perturbative $\protect{\mathbf{c\bar{c}}}$ and 
$\protect{\mathbf{b\bar{b}}}$}
\label{sec:perturbative}

When normalized to $1\%$ probability, the BHPS model predicts that a fraction 
$\int_0^1 [f_{c}(x) + f_{\bar{c}}(x)]\,x\,dx = 0.0057$ of the proton momentum 
is carried by nonperturbative charm. 
The models of Sections~\ref{sec:fivequark}--\ref{sec:MesonBaryon} 
give quite similar values, ranging from $0.0046$ to $0.0073 \,$.

These possible intrinsic momentum fractions can be compared with the standard 
perturbative contributions to the proton momentum, which are shown in 
Fig.~\ref{fig:fig6} as a function of $\mu$.
(These were calculated from the CTEQ6.1 global analysis \cite{cteq6}, 
with uncertainty ranges based on the eigenvector uncertainty 
sets \cite{eigenvector,cteq6}.) 
Note that the $c+\bar{c}$ and $b+\bar{b}$ fractions have been multiplied by 
10 for clarity.  We see that a possible $1\%$ intrinsic charm contribution 
would be rapidly overtaken by perturbatively generated charm, once the 
evolution in $\mu$ has proceeded a short distance above $m_c$.\footnote{%
The rapid rise of $c+\bar{c}$ is likely to be somewhat exaggerated in this 
CTEQ analysis, which uses the standard ``zero mass scheme'' wherein the 
charm quark is treated as a massless parton at scales $\mu > m_c$.}
Gluon splitting similarly generates $b \bar{b}$ pairs rapidly above the scale 
$\mu \! \sim \! m_b$.  Thus intrinsic $c$ and $b$ cannot be expected to add 
significantly to the perturbatively generated $c$ and $b$ for most regions 
of $x$ and $\mu$.

Nevertheless, the intrinsic $c \bar{c}$ component may be very significant 
at large $x$. This is demonstrated by Fig.~\ref{fig:fig7}, which shows the 
probability distributions as a function of $x$, weighted by a factor $x^2$ 
to clarify the large-$x$ region.  The intrinsic component is stronger than 
the perturbative one at $x > 0.3$, even for $\mu$ as large as 
$100 \, \mathrm{GeV}\,$. 
(The intrinsic component will of course also evolve with $\mu$, but that will 
not significantly alter this comparison.)

\figeight

\section{Comparison with light quarks and gluon}
\label{sec:lightquark}

Figure~\ref{fig:fig8} compares the BHPS model, which is representative of 
all the models described here, with the light-quark flavors and gluon 
from CTEQ6.1.
It shows the remarkable result that with the assumption of $1\%$ intrinsic 
charm,  the $c$ and $\bar{c}$ distributions are larger than $\bar{d}$ and 
$\bar{u}$ at large $x$.  This result arises from the $m^2/x$ term in the 
off-shell distance of the heavy-quark states.

Figure~\ref{fig:fig8} also shows that intrinsic $c$ and $\bar{c}$ are much 
smaller than the valence quark and gluon distributions.  This implies that 
the possibility of IC does not significantly affect the light 
quark and gluon distributions.  As a corollary, it negates a pretty 
speculation that IC could be the source of an unexpected 
feature of the CTEQ6.1 PDFs which can be seen in Fig.~\ref{fig:fig8}: the 
gluon distribution is larger than the valence quarks at very large $x$ for 
small $\mu$.  (That feature is not a robust feature of the CTEQ PDFs, since 
it can be made to disappear by a small change in 
the gluon parameterization at $\mu_0$, with an insignificant increase
($\Delta \chi^2 \approx 5$) in the $\chi^2 \approx 2000$ of the global fit.
The feature therefore does not actually require an explanation.)

\section{Conclusion}
\label{sec:conclusion}

\tableI

The light cone ideas used here are at best qualitative and heuristic. 
It is not for example clear whether they should be applied in 
$\overline{\mathrm{MS}}$ or some other 
scheme; or at what scale $\mu_0$.  They should nevertheless be a useful 
guide in the effort to measure intrinsic heavy flavors in the proton.

We have shown that intrinsic charm (IC) will provide the dominant 
contribution to $c$ and $\bar{c}$ at large $x$, if the shape of the IC
distribution is given by the BHPS model and the 
normalization is anywhere near the estimated $1\%$ probability.
All of the other light-cone based models we have examined have roughly the 
same shape in $x$ dependence, and hence they reinforce this result.
Several of the models predict a difference between $c$ and $\bar{c}$, with 
$\bar{c}(x) - c(x) > 0$ at large $x$.
Similar conclusions for the shape apply for intrinsic $b$.

Assuming the $1\%$ probability is approximately correct for IC,
$c$ is much smaller than $u$, $d$, and $g$ at all $x$, so it has 
no appreciable impact on the evolution of other flavors.  Intrinsic $b$ is 
presumably even smaller.

An estimate of $(0.86 \pm 0.60)\%$ was obtained for the IC probability some 
time ago \cite{Hoffmann,HarrisSmithVogt} by re-analysis of $F_2^c$ data in 
deep inelastic muon scattering on iron \cite{EMC}. 
That estimate continues to be cited (see e.g. \cite{LightCone}) as evidence 
for the existence of IC, although it is obviously of limited statistical 
power; and when possible variations in the parton distributions are taken 
into account, the $F_2^c$ data are 
consistent with no IC \cite{SteffensMT,mrstgrv}.  Measurements 
of $F_2^c$ at HERA \cite{F2ccbarbbbar} are also consistent with no IC, but 
those measurements are not at sufficiently large $x$ to have any sensitivity 
to it.  For other experimental indications of IC, see \cite{OtherIC}.

In order to actually measure intrinsic charm or bottom, it will be necessary 
to have data that are directly sensitive to the large-$x$ component. 
Likely candidates are jet 
production with $c$- or $b$-tagging---either inclusively or in association 
with $W$, $Z$, or high-$p_T$ $\gamma$.
It may also be possible to extract useful information from coherent 
diffractive dissociation processes such as $p \to p \, \jpsi$ on a nuclear 
target \cite{Adep}.

For convenience in future work, the model curves (1)--(12) that appear in 
Figs.~1--5 can be adequately represented by a simple parametrization which 
is given in Table~\ref{tab:tableI}.\footnote{%
With the parameters listed for it, this simple parametrization also works 
very well for the BHPS model, though of course the full expression 
(\ref{eq:fq4}) is not inconvenient.}
To create this table, the 
normalization coefficients $A_0$ were chosen to make the momentum fraction
$\int_0^1 f_{c}(x) \, x \, dx$ or $\int_0^1 f_{\bar{c}}(x) \, x \, dx$ 
equal to $0.002857$, the value given by the BHPS 
model when that model is normalized to $1\%$ probability.  This is 
different from the normalization of the curves shown in the Figs.~1--5, 
which was such that each curve corresponded to $1\%$ probability. 
This new normalization is more useful for applications, since it places 
more emphasis on large $x$, where intrinsic charm will be important if 
it is important at all. 
For comparison, the momentum carried by $s$ or $\bar{s}$ at 
$\mu = 1.4 \, \mathrm{GeV}$ is---as it should be---substantially larger 
(by a factor of $4$) than this working estimate of $0.002857$ for $c$ 
or $\bar{c}$. 

\paragraph{Acknowledgements:}
I thank Stan Brodsky for stimulating discussions. 
This research is supported by the National Science Foundation.

\appendix

\section*{Appendix: Light-cone probability distributions from Feynman rules}

This Appendix shows how to derive light-cone probability distributions 
directly from Feynman diagram rules by a thought experiment.\footnote{%
The technique of this thought experiment can in fact be used to calculate
coherent production such as occurs in the ``$\,A^{2/3}\,$'' component of 
$\jpsi$ production on nuclei -- see \cite{Adep}.}

For simplicity, consider a spin 0 particle with mass $m_0$ that couples 
to spin 0 particles with masses $m_1,\dots,m_N$ by a point coupling $i g$,
as illustrated in Fig.~\ref{fig:fig9}(a).  The thought experiment 
consists of scattering this system at very high energy from a target 
that interacts with only one of the constituents, 
as illustrated in Fig.~\ref{fig:fig9}(b). 
This target supplies an 
infinitesimal momentum transfer that puts
the $N$-particle system on mass-shell (diffractive dissociation) 
with a cross section that must be proportional to the probability for that 
system in the original Fock space.

\fignine

Assume that the target provides a constant total cross section $\sigma_0$,
with transverse momentum transfer dependence proportional to
$\exp(-\beta \, \vec{q}_\perp^{\;2})$. 
The elastic differential cross section is 
\begin{equation}
{{d\sigma} \over {d\vec{q}_\perp^{\; 2}}} = 
{{\sigma_0^{\,2}} \over {16\pi}} e^{-\beta \, \vec{q}_\perp^{\; 2}}
\label{eq:app1}
\end{equation}
and hence the integrated elastic cross section is 
\begin{equation}
\sigma_\mathrm{el} = {{\sigma_0^{\, 2}} \over {16 \pi \beta}} \; .
\label{eq:app2}
\end{equation}
By Feynman rules, Fig.~\ref{fig:fig9}(b) gives an amplitude
\begin{equation}
{\mathcal M} = {{2 \, i \, g \, p_N \! \cdot \! q } \over {t - m_N^{\,2}}} \,
          \exp(-\beta \, {\vec{q}_\perp}^{\; 2} /2)
\label{eq:app3}
\end{equation}
and a cross section
\begin{equation}
d\sigma = 
{{4 \, \pi^4} \over {p_0 \! \cdot \! q}} \,
{{d^3q} \over {16 \, \pi^3 \, q^{(0)} }}
\prod_{j=1}^N \! \left( {{d^3 p_j} \over {16 \, \pi^3 \, p_j^{(0)}}} \right) 
\delta^{(4)}(p_0 + k - \sum_{j=1}^N p_j - q) \,
|{\mathcal M}|^2 \, .
\label{eq:app4}
\end{equation}
Assume the gaussian parameter $\beta$ is large, corresponding to 
a large spatial extent of the target in impact parameter.
We can then set $q_\perp = 0$ 
everywhere except in the exponential factor and carry out the 
integral over $q_\perp$.
The Fock space probability density $dP$ can be identified from 
the obvious relation
\begin{equation}
d\sigma = \sigma_\mathrm{el} \times dP \, .
\label{eq:app5}
\end{equation}

Now introduce the light-cone components of all four-momenta,
$p^{(\pm)} = (p^{(0)} \pm p^{(3)})/\sqrt{2}\,$, and define the 
light-cone momentum fractions $x_j = p_j^{(+)}/p_0^{(+)}$.
The components $p_0^{(+)}$ and $q^{(-)}$ are taken to be 
large, with $p_{0 \, \perp} = q_{\perp} = 0$.
The small components are determined by mass-shell 
conditions, e.g., 
$p^2 = m^2$ 
$\Rightarrow$ 
$p^{(-)} = (p_{\perp}^2 + m^2)/(2 p^{(+)})$.
This leads to
\begin{equation}
dP = 
{{1} \over {(16 \, \pi^3)^{N-1}}}
\prod_{j=1}^N \! \left( {{d^2 p_{j {\scriptscriptstyle\perp}} \, dx_j} 
\over {x_j}} \right) \,
\delta^{(2)} (\sum_{j=1}^N p_{j {\scriptscriptstyle\perp}}) \,
\delta(1 - \sum_{j=1}^N x_j) \,
{{x_N^{\,2} \, g^2} \over {(t - m_N^{\,2})^2}} \; .
\label{eq:app6}
\end{equation}

The covariant off-shell distance can be expressed conveniently in the form
\begin{equation}
m_N^{\,2} - t = 
m_N^{\,2} - \left(p_0 - \sum_{j=1}^{N-1} p_j \right)^{\! 2} = 
x_N \, (s - m_0^{\,2}) \; ,
\label{eq:app7}
\end{equation}
where
\begin{equation}
s = 
\sum_{j=1}^{N} 
(p_{j {\scriptscriptstyle\perp}}^{\,2} + m_j^{\,2}) / x_j \; .
\label{eq:app8}
\end{equation}
This leads to the final result 
\begin{equation}
dP = 
{{1} \over {(16 \, \pi^3)^{N-1}}}
\prod_{j=1}^N \! d^2 p_{j {\scriptscriptstyle\perp}} \,
\prod_{j=1}^N \! {{dx_j} \over {x_j}} \,
\delta^{(2)}(\sum_{j=1}^N p_{j {\scriptscriptstyle\perp}}) \,
\delta(1 - \sum_{j=1}^N x_j) \,
{{g^2} \over {(s - m_0^{\,2})^2}} 
\label{eq:app9}
\end{equation}
for the point-coupling model.
Note that this result is completely symmetric in the particles $1,\dots,N$ 
as it should be---it does not depend on which particle was singled out to 
scatter in the thought experiment used to derive it. 
It is straightforward to include more complicated vertices and factors due 
to spin using this Feynman diagram approach.  When that is done, the 
result \textit{can} depend on which particle is assumed to scatter, 
but the ambiguity vanishes at the pole at $s = m_0^{\,2}$.
Unitarity effects that keep the total probability equal to $1$ could also be 
included.

In this simple point-coupling model, high-mass Fock states are suppressed 
only by the ``old-fashioned perturbation theory energy denominator'' factor
$(s - m_0^{\,2})^{-2}\,$.  To make the model more realistic, there must be 
a further suppression of high mass states associated with wave function 
effects---if only to make the integrated probability finite.  It is natural 
to suppose that the additional suppression is a 
function of $s$.\footnote{%
It is \textit{not} natural to assume that the 
suppression is a function of the covariant variable $t$, as is done in some 
``meson cloud'' models \cite{mesoncloud1}, since that assumption spoils 
the independence on which particle is taken to be off shell.  A related 
argument leading to this conclusion is given in more recent meson cloud 
work \cite{mesoncloud2}.}

When a wave function factor $[F(s)]^2$ is included in Eq.~(\ref{eq:app9}), the 
transverse momentum integrals can be carried out by inserting the identity 
$1 = \int \delta(\sum_j
(p_{j {\scriptscriptstyle\perp}}^{\,2} + m_j^{\,2}) / x_j - s) \, ds$, 
and then Fourier transforming this delta 
function and the transverse momentum conserving one.  The result is 
\begin{equation}
dP = 
{{g^2} \over {(16 \, \pi^2)^{N-1} \, (N-2)!}}
\prod_{j=1}^N \! dx_j \,
\delta(1 - \sum_{j=1}^N x_j) \,
\int_{s_0}^\infty 
{{(s-s_0)^{N-2} \, [F(s)]^2 \, ds} \over {(s - m_0^{\,2})^2}}
\label{eq:app10}
\end{equation}
where $s_0 = \sum_{j=1}^{N} m_j^{\,2} / x_j \,$.  Note that $[F(s)]^2$ must 
go to zero faster than $1/s^{N-3}$ as $s \to \infty$ to make the 
integrated probability converge.


\begin{thebibliography}{99}

\bibitem{lattice}
J.~W.~Negele {\it et al.},
Nucl.\ Phys.\ Proc.\ Suppl.\  {\bf 128}, 170 (2004)
[hep-lat/0404005];
W.~Schroers,
Nucl.\ Phys.\ A {\bf 755}, 333 (2005)
[hep-ph/0501156].

\bibitem{cteq6} 
J.~Pumplin, D.~R.~Stump, J.~Huston, H.~L.~Lai, P.~Nadolsky and W.~K.~Tung,
JHEP {\bf 0207}, 012 (2002)
[hep-ph/0201195];
D.~Stump, J.~Huston, J.~Pumplin, W.~K.~Tung, H.~L.~Lai, S.~Kuhlmann 
and J.~F.~Owens,
JHEP {\bf 0310}, 046 (2003)
[hep-ph/0303013].

\bibitem{mrst} 
R.~S.~Thorne, A.~D.~Martin, W.~J.~Stirling and R.~G.~Roberts,
[hep-ph/0407311].

\bibitem{F2ccbarbbbar}
A.~Aktas {\it et al.}  [H1 Collaboration],
Eur.\ Phys.\ J.\ C {\bf 40}, 349 (2005)
[hep-ex/0411046]; 
A.~Aktas {\it et al.}  [H1 Collaboration],
[hep-ex/0507081].

\bibitem{LightCone} 
S.~J.~Brodsky,
``Light-front QCD,''
[hep-ph/0412101];
S.~J.~Brodsky,
Few Body Syst.\  {\bf 36}, 35 (2005)
[hep-ph/0411056].

\bibitem{BrodskyHoyer}
S.~J.~Brodsky, P.~Hoyer, C.~Peterson and N.~Sakai,
Phys.\ Lett.\ B {\bf 93}, 451 (1980).

\bibitem{Brodsky}
S.~J.~Brodsky, C.~Peterson and N.~Sakai,
Phys.\ Rev.\ D {\bf 23}, 2745 (1981);
R.~Vogt,
Prog.\ Part.\ Nucl.\ Phys.\  {\bf 45}, S105 (2000)
[hep-ph/0011298];
T.~Gutierrez and R.~Vogt,
Nucl.\ Phys.\ B {\bf 539}, 189 (1999)
[hep-ph/9808213];
G.~Ingelman and M.~Thunman,
Z.\ Phys.\ C {\bf 73}, 505 (1997)
[hep-ph/9604289];
J.~Alwall and G.~Ingelman,
Phys.\ Rev.\ D {\bf 71}, 094015 (2005)
[hep-ph/0503099];
J.~Alwall,
[hep-ph/0508126].

\bibitem{Polyakov}
M.~Franz, .~V.~Polyakov and K.~Goeke,
Phys.\ Rev.\ D {\bf 62}, 074024 (2000)
[hep-ph/0002240].

\bibitem{Navarra}
F.~S.~Navarra, M.~Nielsen, C.~A.~A.~Nunes and M.~Teixeira,
Phys.\ Rev.\ D {\bf 54}, 842 (1996)
[hep-ph/9504388];
S.~Paiva, M.~Nielsen, F.~S.~Navarra, F.~O.~Duraes and L.~L.~Barz,
Mod.\ Phys.\ Lett.\ A {\bf 13}, 2715 (1998)
[hep-ph/9610310].

\bibitem{Melnitchoukccbar}
W.~Melnitchouk and A.~W.~Thomas,
Phys.\ Lett.\ B {\bf 414}, 134 (1997)
[hep-ph/9707387].

\bibitem{SteffensMT}
F.~M.~Steffens, W.~Melnitchouk and A.~W.~Thomas,
Eur.\ Phys.\ J.\ C {\bf 11}, 673 (1999)
[hep-ph/9903441].

\bibitem{strangecloud}
A.~I.~Signal and A.~W.~Thomas,
Phys.\ Lett.\ B {\bf 191}, 205 (1987);
W.~Koepf, L.~L.~Frankfurt and M.~Strikman,
Phys.\ Rev.\ D {\bf 53}, 2586 (1996)
[hep-ph/9507218];
S.~J.~Brodsky and B.~Q.~Ma,
Phys.\ Lett.\ B {\bf 381}, 317 (1996)
[hep-ph/9604393].

\bibitem{BagModel}
J.~F.~Donoghue and E.~Golowich,
Phys.\ Rev.\ D {\bf 15}, 3421 (1977).

\bibitem{Song}
X.~T.~Song,
Phys.\ Rev.\ D {\bf 65}, 114022 (2002)
[hep-ph/0111129].

\bibitem{BrodskyMa}
S.~J.~Brodsky and B.~Q.~Ma,
Phys.\ Lett.\ B {\bf 381}, 317 (1996)
[hep-ph/9604393].

\bibitem{eigenvector} 
J.~Pumplin, D.~R.~Stump and W.~K.~Tung,
Phys.\ Rev.\ D {\bf 65}, 014011 (2002)
[hep-ph/0008191].

\bibitem{Hoffmann}
E.~Hoffmann and R.~Moore,
Z.\ Phys.\ C {\bf 20}, 71 (1983).

\bibitem{HarrisSmithVogt} 
B.~W.~Harris, J.~Smith and R.~Vogt,
Nucl.\ Phys.\ B {\bf 461}, 181 (1996)
[hep-ph/9508403].

\bibitem{EMC}
J.~J.~Aubert {\it et al.}  [European Muon Collaboration],
Nucl.\ Phys.\ B {\bf 213}, 31 (1983).

\bibitem{OtherIC}
R.~Vogt, S.~J.~Brodsky and P.~Hoyer,
Nucl.\ Phys.\ B {\bf 360}, 67 (1991);
R.~Vogt, S.~J.~Brodsky and P.~Hoyer,
Nucl.\ Phys.\ B {\bf 383}, 643 (1992);
R.~Vogt and S.~J.~Brodsky,
Phys.\ Lett.\ B {\bf 349}, 569 (1995)
[hep-ph/9503206];
R.~Vogt and S.~J.~Brodsky,
Nucl.\ Phys.\ B {\bf 478}, 311 (1996)
[hep-ph/9512300].


\bibitem{mrstgrv}
A.~D.~Martin, R.~G.~Roberts, W.~J.~Stirling and R.~S.~Thorne,
Eur.\ Phys.\ J.\ C {\bf 4}, 463 (1998)
[hep-ph/9803445];
M.~Gluck, E.~Reya and A.~Vogt,
Eur.\ Phys.\ J.\ C {\bf 5}, 461 (1998)
[hep-ph/9806404].

\bibitem{Adep} 
J.~Badier {\it et al.}  [NA3 Collaboration],
Z.\ Phys.\ C {\bf 20}, 101 (1983);
S.~J.~Brodsky and P.~Hoyer,
Phys.\ Rev.\ Lett.\  {\bf 63}, 1566 (1989).

\bibitem{mesoncloud1} 
A.~I.~Signal, A.~W.~Schreiber and A.~W.~Thomas,
Mod.\ Phys.\ Lett.\ A {\bf 6}, 271 (1991);
S.~Kumano,
Phys.\ Rev.\ D {\bf 43}, 59 (1991);
S.~Kumano,
Phys.\ Rev.\ D {\bf 43}, 3067 (1991).

\bibitem{mesoncloud2} 
W.~Melnitchouk, J.~Speth and A.~W.~Thomas,
Phys.\ Rev.\ D {\bf 59}, 014033 (1999)
[hep-ph/9806255].

\end{thebibliography}
\end{document}